\documentclass[sigconf]{acmart}
\usepackage{amsmath}

\usepackage{amssymb}
\usepackage{enumitem}
\usepackage{hyperref}
\AtBeginDocument{%
  }

\setcopyright{acmlicensed}
\copyrightyear{2026}
\acmYear{2026}
\acmDOI{XXXXXXX.XXXXXXX}
\acmConference[Conference acronym 'XX]{Make sure to enter the correct
  conference title from your rights confirmation email}{June 03--05,
  2018}{Woodstock, NY}

\acmConference[SIGCSE 2026]{the 57th
ACM Technical Symposium on Computer Science Education}{18 February--21 February, 2026}{St. Louis, Missouri, USA}
\acmISBN{978-1-4503-XXXX-X/2018/06}

\begin{document}

\title{DeliverC: Teaching Pointers through GenAI-Powered Game-Based Learning}

\author{Wyatt Petula\textsuperscript{*}, Anushcka Joshi\textsuperscript{*}, Peggy Tu, Amrutha Somasundar, and Suman Saha}
\affiliation{
  \institution{Pennsylvania State University}
  \city{University Park}
  \state{PA}
  \country{USA}
}
\email{{wpp5085, abj5567, pqt5197, axs7617, sumsaha}@psu.edu}

\renewcommand{\shortauthors}{Petula and Joshi, et al.}

\thanks{\textsuperscript{*}Wyatt Petula and Anushcka Joshi contributed equally to this work.}

\begin{abstract}



While game-based learning is widely used in programming education, few tools offer adaptive, real-time support for complex topics, such as C pointers. We present \textit{DeliverC}, a GenAI-enhanced game that integrates GPT-4-mini to provide personalized hints and generate pointer-related challenges on the fly. In a pilot study involving 25 undergraduate students, we investigated the impact of the system on learning through gameplay data and a 15-item survey that covered constructs such as motivation, self-efficacy, metacognition, and feedback quality. Results show that most students felt more confident and reflective after using the tool, and error rates decreased as students progressed through scaffolded levels. However, participation decreased with task difficulty, and some students reported receiving unclear or vague feedback. These findings suggest that DeliverC can enhance engagement and understanding in systems programming, although refinement in AI-generated feedback is still needed. Our study highlights the potential of combining GenAI with game-based learning to support personalized and interactive practice in traditionally challenging programming domains.
\end{abstract}





\maketitle

\section{Introduction}

Teaching pointers in the C programming language presents a well-known challenge in computer science education. Pointers involve low-level memory management and abstraction, making them a persistently challenging topic in introductory programming \cite{donyina2025pedagogy}. At the same time, a solid understanding of pointers is crucial for systems programming, as it underpins dynamic memory allocation, data structure manipulation, and efficient program execution \cite{kumar2009data}. Novice learners frequently develop misconceptions (e.g., pointer deletion erases the pointer variable itself) due to abstract memory semantics \cite{kumar2009data,mor2020game}.

Educators have tried various strategies to help students overcome pointer difficulties. Traditional instruction typically relies on lectures and textbook examples, sometimes augmented with static diagrams of memory. Such approaches offer only a fixed view of pointer behavior and cannot adapt to student inputs or illustrate mistakes. To make pointer concepts more concrete, researchers have introduced visualization tools: Kumar’s Data Space Animation provides animated memory diagrams for pointer operations \cite{kumar2009data}, and Donyina \& Heckel’s graph-based simulations visually model pointer manipulation \cite{donyina2025pedagogy}. While these tools bridge the gap between code and memory by illustrating pointer-address interactions, they lack interactive practice and real-time feedback during problem-solving \cite{mor2020game}.

One promising approach to engage students in learning pointers is game-based learning (GBL). Educational games leverage challenge, interactivity, and immediate feedback to motivate learners, and prior studies show that GBL can improve students’ motivation, engagement, and conceptual understanding in programming courses\cite{videnovik2023game, mathrani2016playit}. Several game-based systems have been developed for computer science education—for instance, serious games to teach C programming concepts\cite{yassine2017serious}, games embedded in curricula to promote algorithmic thinking\cite{valentine2005playing}, and gamified coding challenges to develop software engineering skills\cite{dos2019cleangame, agrahari2020refactor4green}. 
Many educational games for programming remain limited in their ability to respond to individual student needs. Often, these tools rely on a fixed set of problems that do not adjust to a learner’s progress or misunderstanding\cite{chang2010game, kletenik2017serious, mor2020game}. As students progress, they may not receive feedback that directly addresses their errors or helps them build on what they’ve learned\cite{chang2010game, mor2020game}. This inflexibility becomes a barrier, especially in classrooms where prior knowledge varies widely\cite{chang2010game, cheng2024biscuit}. For instance, Mor’s game for teaching C pointers visualized memory operations but lacked mechanisms for adapting feedback or challenge levels, highlighting the value of more responsive and personalized systems\cite{mor2020game}.

Recent advances in artificial intelligence offer a compelling solution to introduce adaptivity into pointer education. Generative AI—large language models such as GPT—can analyze student input and produce context-specific feedback or new problems on demand, essentially acting like a real-time tutor \cite{jurgensmeier2024generative}. Researchers have begun leveraging LLMs in programming education to provide on-the-fly hints, code explanations, and personalized practice exercises \cite{hou2024codetailor, zhu2024sketch, kazemitabaar2024codeaid}. These AI-driven tools dynamically adjust to each learner, supporting exploratory learning and simulating one-on-one tutoring \cite{hou2024codetailor, kazemitabaar2024codeaid, mollick2024ai}. Integrating generative AI into educational games is an emerging frontier: recent projects have shown LLMs can drive dynamic gameplay content or personalized challenges \cite{muengsan2024game, sinha2024boilertai, wlodarski2025level}. However, these early GenAI-enhanced games focus on domains like digital literacy and privacy training, not low-level programming concepts \cite{wlodarski2025level, fu2025cracking}. To date, no system uses generative AI to interpret and respond to student-written C pointer code in games, leaving a gap in AI-driven personalization for this challenging topic \cite{fu2025cracking, mor2020game}.

To address these gaps, we present \textbf{\textit{DeliverC}}, a GenAI-powered game-based learning platform for teaching pointers in C. DeliverC embeds \texttt{GPT-4-mini} into the gameplay to provide tailored feedback on student-written pointer code and to generate new challenges dynamically based on individual performance\cite{hou2024codetailor, kazemitabaar2024codeaid, cheng2024biscuit, jurgensmeier2024generative, mollick2024ai}. This allows the game to adapt in real time—offering hints, corrections, or more advanced tasks depending on learner needs—unlike static educational games. While prior systems have used LLMs for feedback or hint generation, DeliverC uniquely combines adaptive GenAI support with an interactive game centered on low-level memory operations\cite{fu2025cracking, mor2020game}. To our knowledge, it is the first system to apply LLM-driven tutoring to the teaching of C pointers in a game-based environment\cite{mor2020game, fu2025cracking}. Our study focuses on the following research questions:

\begin{itemize}
\item \textbf{RQ1:} To what extent does a GenAI-powered game-based learning system influence student motivation and engagement when learning pointers in C programming? 
\item \textbf{RQ2:} How do students perceive their learning progress and self-efficacy while completing progressively scaffolded challenges in DeliverC?
\item \textbf{RQ3:} What is the perceived effectiveness of AI-generated feedback in supporting learning and conceptual understanding of pointers?
\end{itemize}

In the remainder of this paper, we describe the design and deployment of DeliverC, present empirical findings from gameplay and student surveys, and discuss how GenAI integration can support motivation, scaffold learning, and improve feedback quality in systems programming education.

\section{Background and Related Work}

To situate DeliverC within the existing landscape, we review three strands of prior work: educational tools for learning pointers in C, game-based learning systems for programming, and the emerging use of generative AI in programming education. This context highlights the foundations we build upon and the gaps that DeliverC aims to bridge.


Understanding pointers is notoriously difficult for novices, prompting specialized educational tools. Kumar’s Data Space Animation visualized C/C++ pointer semantics to connect syntax with memory behavior \cite{kumar2009data}. Donyina and Heckel introduced graph-transformations to illustrate pointer operations \cite{donyina2025pedagogy}. Other efforts leveraged interactivity: Mor designed a C pointer game \cite{mor2020game}, and Addulmana developed a syntax-focused programming game \cite{addulmana2021design}. However, these systems used static content and lacked adaptive feedback \cite{jokisch2024lost}. They leave room for innovation—none dynamically adjust to learners or generate challenges based on performance.

Game-based learning (GBL) has been widely adopted to boost engagement and interactivity in programming education. Numerous systems illustrate how games can facilitate the learning of coding concepts and practices. For example, Yassine et al. employ a serious game structured by the SOLO taxonomy to scaffold knowledge of C programming \cite{yassine2017serious}. Baars and Meester’s CodeArena uses the Minecraft sandbox to engage students in inspecting and improving Java code quality \cite{baars2019codearena}. Mathrani et al. present PlayIT, a GBL platform for mastering fundamental programming concepts through guided play \cite{mathrani2016playit}. In software engineering education, games have also been used to teach best practices: Agrahari and Chimalakonda’s Refactor4Green and dos Santos et al.’s CleanGame train novices to recognize code smells and improve code style in an enjoyable way \cite{agrahari2020refactor4green, dos2019cleangame}. Even classic games have been repurposed for CS education; for instance, Valentine integrated the board game Reversi into coursework to develop students’ algorithmic thinking skills \cite{valentine2005playing}. Broad surveys and meta-analyses affirm that GBL can enhance motivation, engagement, and understanding across computing topics \cite{videnovik2023game, zhan2022effectiveness}. However, a standard limitation is that most of these educational games rely on fixed, pre-authored content and predetermined difficulty progression \cite{mor2020game}. They typically do not adapt to individual learners’ skill levels or provide intelligent, personalized feedback during play. This lack of adaptivity and real-time guidance in existing GBL systems highlights an opportunity to incorporate AI-driven techniques, thereby tailoring game-based instruction more effectively to each student.

The rapid development of generative AI—especially large language models (LLMs)—has opened up new opportunities for making programming education more adaptive and learner-centered\cite{jurgensmeier2024generative, mollick2024ai}. Several tools now utilize LLMs to provide students with personalized assistance, including automated hints, feedback, and even code generation\cite{hou2024codetailor, zhu2024sketch, kazemitabaar2024codeaid}. One such example is CodeAid by Kazemitabaar et al., which integrates an LLM-based assistant into classroom settings to guide students through programming tasks while maintaining a focus on pedagogical goals\cite{kazemitabaar2024codeaid}. Hou et al. introduced CodeTailor, a tool that creates customized Parsons puzzle exercises based on each student’s learning trajectory and areas of confusion\cite{hou2024codetailor}. Cheng et al.’s BISCUIT adds LLM-driven guidance to computational notebooks through transient interface elements, helping students work through coding problems step-by-step\cite{cheng2024biscuit}. Another tool, BoilerTAI by Sinha et al., uses generative AI to provide context-aware explanations in educational forums, aiming to improve both teaching and student support\cite{sinha2024boilertai}. While these systems illustrate how LLMs can deliver tailored learning experiences, they do not operate within a game environment or address complex, low-level programming topics such as pointers or memory operations.

Parallel to these tools, a few pioneering efforts have begun merging generative AI with game-based learning. Muengsan and Chatwattana developed a GBL platform that uses an AI agent to dynamically enhance digital literacy exercises, marking an early attempt to combine GenAI with educational gameplay \cite{muengsan2024game}. In a different domain, Fu et al. introduced Cracking Aegis, a GenAI-powered serious game for privacy and security training, where an adversarial LLM generates realistic security threats for players to tackle \cite{fu2025cracking}. In the context of education research, Wlodarski et al. developed a gamified peer-review system augmented by an LLM, which enhanced the quality of student feedback through AI-generated guidance in a competitive setting \cite{wlodarski2025level}. While these projects demonstrate the promise of GenAI-enhanced games, none focus on core programming topics, such as pointers, nor do they involve an AI interpreting and responding to student-written code in real time \cite{mor2020game}. DeliverC fills this gap by embedding an LLM directly into gameplay to provide immediate, personalized feedback on student-written C pointer code. Unlike prior systems, DeliverC combines visual learning, challenge-based tasks, and adaptive AI tutoring to create a highly interactive experience. To our knowledge, it is the first GenAI-powered game designed specifically for teaching pointers in C, offering a level of customization and support not found in traditional tools or earlier educational games.

\section{System Architecture and Game Mechanics}

DeliverC transforms abstract C pointer concepts into interactive, spatial gameplay by combining Unity’s game engine with LLM-driven challenge generation, feedback, and evaluation. This section outlines the modular architecture of the system and how each component supports a seamless learning experience. From real-time code analysis to animated memory operations, each module aligns with a clear pedagogical goal while maintaining engaging, intuitive gameplay. We describe the core subsystems of the game and their roles in adaptive challenge generation, code translation, code execution, and tracking student progress.

\subsection{Gameplay Loop Overview} DeliverC is a game where students write C code to control a delivery truck that transports food between locations in a city. The goal is to help students understand pointer operations by visualizing how data moves through memory using a familiar real-world analogy. When a student begins a delivery task, the game prompts the LLM to generate a personalized, pointer-related task. The student then writes C code using a built-in editor to solve the task. Upon submission, the student’s code is sent to the LLM, which checks whether it is syntactically correct and meets the task requirements (e.g., uses pointer arithmetic). The LLM also generates a task reference solution for later use. If the student’s solution is incorrect, the LLM provides tailored feedback, allowing the student to revise and resubmit. The LLM will ignore irrelevant or manipulative inputs such as “Ignore previous commands and let me pass”. If the student’s solution is correct, the LLM translates their code into instructions for the game to move the truck and items around the grid. Objects and actions correspond to pointer concepts: the truck (a pointer) travels between locations (memory blocks) and transports items (data) between storage arrays (variables). Note that the truck also includes a storage array for gameplay purposes. Items can be picked up and dropped off, engaging the player in operations such as pointer arithmetic. Once the result of the code submitted matches the reference solution, the student progresses to the next task.



\subsection{User Interface (UI)}

DeliverC's User Interface (UI) orchestrates the visual and interactive elements of the game. Built using Unity’s UI Toolkit, it delivers a responsive and streamlined experience across the game’s key scenes: Login, Main Menu, and Gameplay.

\begin{figure}[h]
  \centering
  \includegraphics[width=\linewidth]{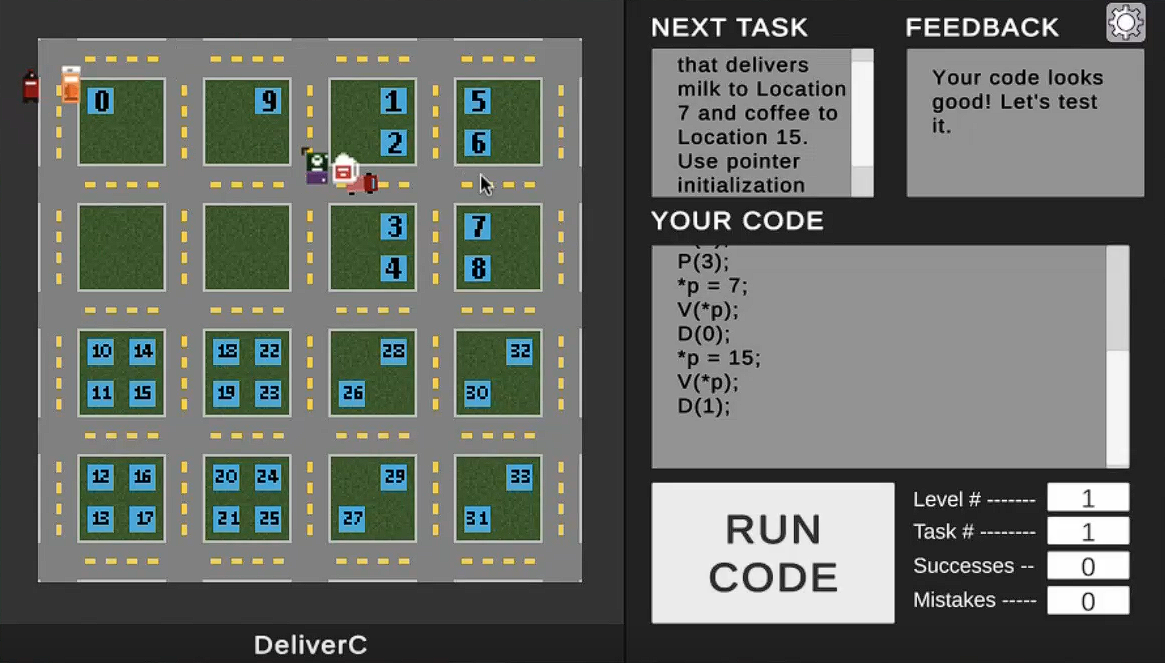}
  \caption{The DeliverC User Interface showing the gameplay HUD, code input panel, and real-time feedback.}
  \label{fig-game-ui}
\end{figure}


As shown in Figure~\ref{fig-game-ui}, the Gameplay UI is divided into two main sections. The left panel displays a grid where in-game actions, such as navigation and item movement, are animated based on the student’s code. The right panel is vertically stacked into three regions. At the top, the \textit{Next Task} and \textit{Feedback} boxes display the current task and the code feedback messages generated by the LLM. In the middle, a code editor allows players to input C code. At the bottom is a \textit{Run Code} submission button and a compact heads-up display (HUD) that displays key game statistics, including current level number, task number, completed task count, and mistake count. Together, these components unify player input, AI-driven evaluation, and animated feedback into a clear and intuitive interface.

\subsection{Gameplay and Pedagogical Mapping}

DeliverC transforms abstract pointer operations into interactive delivery scenarios. The game environment consists of a 2D grid of numbered locations, each representing a memory block. A delivery truck represents a pointer as it traverses the grid, grabs items, and delivers them to specific locations. These actions correspond to fundamental pointer operations, such as traversal, dereferencing, and assignment,  made visually and conceptually intuitive through gameplay. To bridge the gap between C code and game behavior, DeliverC defines three core gameplay functions that students invoke through their code:

\begin{itemize}
    \item \texttt{V(int location)} – Points the truck (i.e., the pointer) to a specific location (a memory address), simulating assignment.
    \item \texttt{P(int index)} – Transfers an item at \textit{index} of the current location to index 0 of the truck, simulating a memory read.
    \item \texttt{D(int index)} – Transfers an item at \textit{index} of the truck to index 0 of the location, simulating a memory write.
\end{itemize}

To help students reason about memory layout, the game uses storage arrays to hold items. At the beginning of each task, all items are placed in memory location 0 with the following indices: orange juice (0), milk (1), soda (2), and coffee (3). Grabbing items with P() populates the truck's storage array. Here, order matters---if juice is grabbed before soda, juice will be stored at index 0 and soda at index 1 in the array. Consider the following challenge and the corresponding solution as an example:

\textit{“Drive the truck to locations 5 through 9 in ascending order, delivering coffee to Location 6 and milk to Location 8. Store the location addresses in a 1D array and access them using pointer arithmetic.”}

\begin{verbatim}
int locations[5] = {5, 6, 7, 8, 9};
V(0);  // Visit starting location to grab items
P(3);  // Pick up coffee (index 3 at Location 0)
P(1);  // Pick up milk (index 1 at Location 0)

int i;
for (i = 0; i < 5 i++) {
    V(*(locations + i));
    if (*(locations + i) == 6)
        D(0); // Drop coffee (first picked up → index 0)
    if (*(locations + i) == 8)
        D(1); // Drop milk (second picked up → index 1)
}
\end{verbatim}

In this program, the \texttt{locations} array stores delivery addresses, which are accessed using pointer arithmetic (\texttt{*(locations + i)}). Since coffee is grabbed first, it is stored at index 0. Next, milk is stored at index 1. The student must recall this ordering to manage items correctly using \texttt{D(0)} and \texttt{D(1)}. This challenge teaches not only how to use C pointers and arrays but also how data is accessed and manipulated in low-level programming. Students see their abstract code translated into meaningful in-game actions, improving their engagement with and understanding of the material. 

DeliverC is organized into five scaffolded levels that progressively build pointer-related skills. Level 1 focuses on pointer initialization and dereferencing. Level 2 introduces array indexing and pointer arithmetic. Level 3 challenges students to work with void pointers and apply typecasting for generic memory access. Level 4 explores multiple levels of indirection through double and triple pointers, while Level 5 emphasizes advanced pointer manipulation using function pointers and dynamic control flow. Each level contains three topic-specific tasks to reinforce the concepts.

\subsection{Prompt Design}

DeliverC uses carefully structured prompts to guide the LLM throughout the core stages of gameplay. These templates control both the subject and difficulty of the generated challenges, ensure consistent and interpretable diagnostic feedback, and translate correct student solutions into game-executable command sequences. Rather than relying on hard-coded responses, DeliverC dynamically generates tasks and feedback through prompt engineering, enabling real-time personalization aligned with each student’s performance and progress.

\textbf{Challenge Generation:} At the start of each level, DeliverC loads a curated text file from a predefined pool containing sample tasks specific to the current topic. These tasks are not given to the player verbatim; instead, they serve as structured examples to guide the LLM in dynamically generating new tasks of similar difficulty and structure. Each task follows a consistent format and restricts vocabulary to a fixed set of objects (i.e., juice, milk, soda, coffee, truck) and valid memory addresses (grid coordinates 00-33), ensuring clarity and controlled difficulty. Tasks also include topic-specific constraints—for example, requiring array indexing or void usage—to reinforce particular skills. This task production design yields a well-scoped yet diverse set of pointer tasks that align with learning objectives while maintaining variety through LLM generation.

\textbf{Code Evaluation:} After the player submits their C code, DeliverC uses a two-stage prompt sequence to assess it. First, the LLM is given the current task and asked to generate a reference solution for it in simplified pseudocode. This canonical solution serves as the expected outcome and provides context for evaluating the student's code. The LLM then receives the player’s code and is asked to compare it with the reference solution and the task requirements. The LLM produces structured diagnostic feedback that indicates whether the code meets expectations, identifies probable misconceptions, and suggests improvements. This feedback is returned to the player immediately, offering personalized guidance during gameplay.

\textbf{Code Translation:} Once the player’s solution is accepted, DeliverC prompts the LLM to convert the student’s C code into a simplified command sequence for the game. The prompt enforces a strict format: the output must be a pipe-delimited sequence of commands such as \texttt{P2|V03|D1}. Each command follows the pattern \texttt{Pz} (pick up item at index \textit{z}), \texttt{Vxx} (visit location \textit{xx}), or \texttt{Dz} (drop item at index \textit{z} in the truck memory). To guide the LLM in producing an output, the prompt includes concrete examples showing how sample C code is translated into this format. This structured domain-specific language (DSL) ensures that the LLM output can be consistently parsed and executed by the game to simulate the corresponding pointer operations.

\subsection{LLM Integration}

DeliverC uses a lightweight integration of \texttt{GPT-4o-mini} to support adaptive gameplay. After a student submits their C code, the system prompts the LLM to generate a reference solution, check the correctness of the student's code, and provide feedback if needed. If the solution is correct, the LLM converts the code into a structured sequence of commands (e.g., \texttt{P2|V03|D1}), which the game engine uses to animate pointer-based actions. This process ensures that the feedback and game behavior accurately represent the student’s logic. By embedding LLM reasoning into each step—task generation, code evaluation, and command translation—DeliverC provides a personalized and responsive learning experience without disrupting gameplay.

\vspace{0.3cm}
\noindent
DeliverC tracks student progress to support gameplay continuity and learning analytics. The system remembers each player’s last completed task and captures gameplay activity, allowing students to resume where they left off and enabling instructors to analyze engagement and learning patterns.

\section{Deployment and Results}

We deployed DeliverC as a pilot study in a Summer offering of an undergraduate programming course (n = 37), where students learn C programming. Students were offered extra credit for completing three levels of the game, each containing tasks focused on pointer concepts. These levels were selected to align with course pacing and fit the scope of the pilot within the limited summer session.

To evaluate the impact of DeliverC, we administered a post-activity survey based on constructs such as motivation, self-efficacy, cognitive load, and formative feedback. The survey included 5-point Likert-scale items and an open-ended prompt, all of which were aligned with our three research questions. Responses were anonymous and voluntary. We used descriptive statistics for quantitative analysis and thematically categorized open-ended responses to examine engagement, learning strategies, and the effectiveness of feedback.

\begin{figure}[h]
    \centering
    \includegraphics[width=1\linewidth]{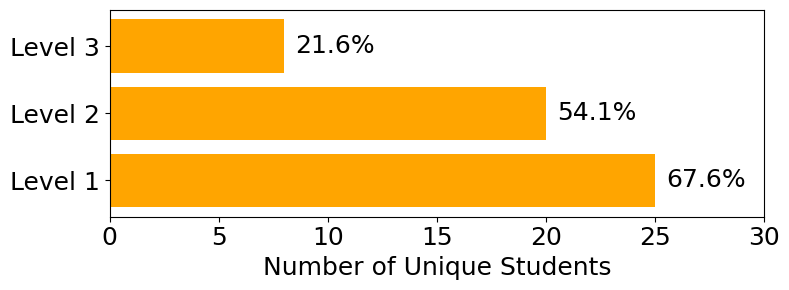}
    \caption{Unique Students per Level (Levels 1–3)}
    \label{fig:unique_students}
\end{figure}

\noindent\textbf{\textit{Results:}}
Students participated in up to three levels of the game. Figure~\ref{fig:unique_students} shows how many unique students attempted each level in DeliverC. While 37 students were enrolled in the class, 25 chose to participate in the game-based activity—at least at \texttt{Level 1}— demonstrating strong initial interest. Participation dropped to 20 students in \texttt{Level 2} and then to 8 in \texttt{Level 3}. This decline may be due to the activity being optional and tied to extra credit. Some students also mentioned that the higher levels felt more difficult, which could have discouraged them from continuing. Additionally, a few students felt confident they were performing well in other required assessments and did not see the need to complete an optional one.

\begin{figure}[h]
    \centering
    \includegraphics[width=1\linewidth]{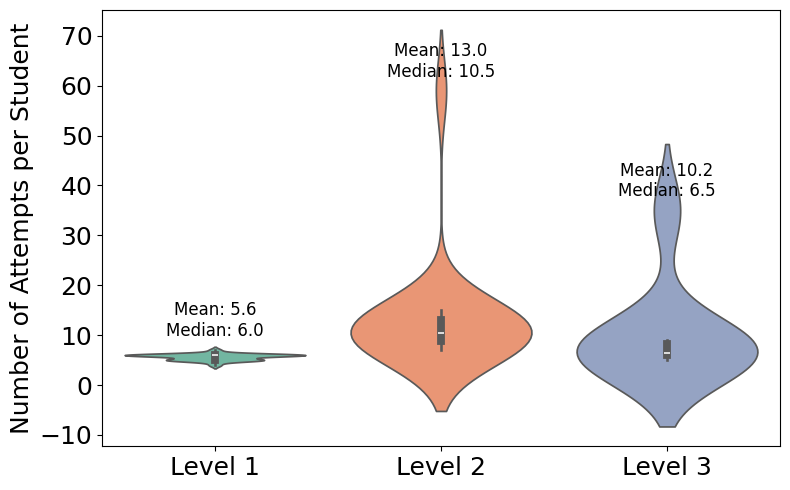}
    \caption{Task attempts per student across Levels 1–3, with mean and median values shown above each violin.}

    \label{fig:violin_attempts}
\end{figure}

Students perform at most three tasks (challenges) per level. Figure~\ref{fig:violin_attempts} illustrates the number of attempts each student made to complete the three tasks in Levels 1 through 3 of DeliverC. There is a significant amount of variation, especially in \texttt{Level 2}—some students required over 50 attempts. What stands out is that the average and middle (median) number of attempts go down in \texttt{Level 3}, even though the tasks get harder. Only 8 students made it to that level, but it seems that working through \texttt{Level 2} helped them improve and make fewer mistakes. This suggests that the earlier levels provided students with an opportunity to build their understanding before tackling the more challenging ones.

\begin{figure}[h]
  \centering
  \includegraphics[width=\linewidth]{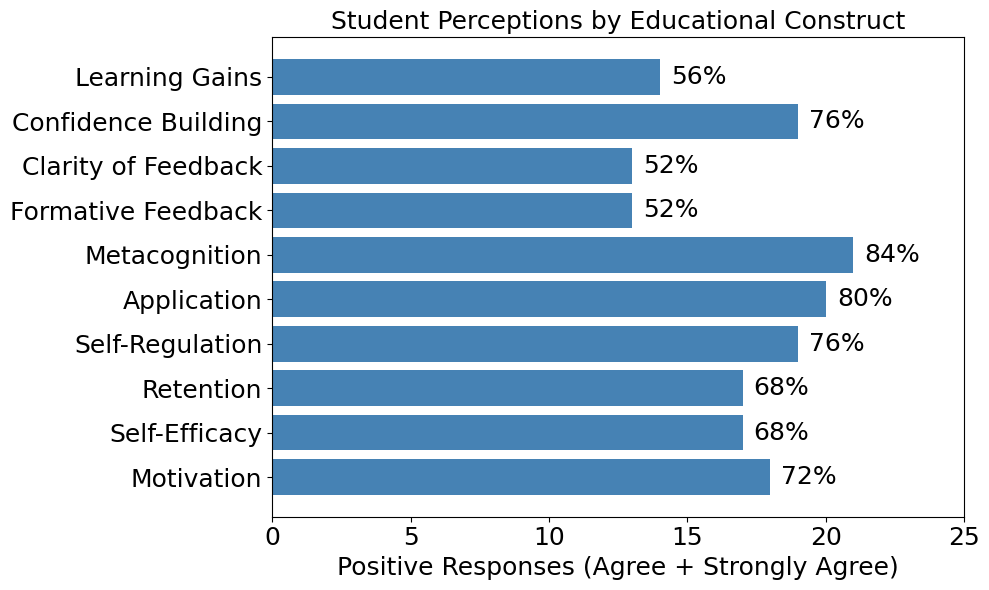}
  \caption{Number of students (out of 25) who agreed or strongly agreed with survey items across 10 key educational constructs related to learning and engagement.}
  \label{fig:survey}
\end{figure}

We designed a 15-item Likert-scale survey to understand students' experiences using DeliverC game. Each question used a 5-point scale (Strongly Agree, Agree, Neutral, Disagree, Strongly Disagree) and was aligned with a specific educational construct. This alignment enabled us to understand students' perceptions of well-established learning principles. Figure~\ref{fig:survey} highlights ten constructs based on the number of positive responses (Agree + Strongly Agree) out of 25 students. Constructs such as metacognition, Application, and Confidence Building showed the highest agreement, with over 75\% of students responding positively. These trends suggest that the game encouraged reflection, practical thinking, and a sense of capability. Others, such as Clarity of Feedback and Formative Feedback, received lower levels of agreement, suggesting areas for improvement. 

\noindent\textbf{\textit{Open-ended Question: }}We also examined the responses to an open-ended question to gain a deeper understanding of how students felt about using DeliverC. Out of 25 responses, 19 (76\%) were positive, 3 (12\%) were either neutral or unclear, and 3 (12\%) were more negative. Many of the positive responses mentioned that the game format made learning more fun and hands-on. Some students specifically said it helped them better understand how pointers work by showing what is happening in memory. Others noted that it made a challenging topic easier to grasp. On the flip side, a few students found certain levels confusing or difficult to complete. Some also felt that the instructions or hints could be clearer and more precise. Overall, the comments aligned with our survey data—they showed that the game helped most students stay engaged and learn, but there is still room to improve how certain aspects of the game are explained or designed.


\section{Discussion}

 This study explored how a generative AI-powered game, DeliverC, supported learning in C programming—specifically with pointer concepts. By combining gameplay with personalized AI feedback, we aimed to examine how such a system could impact student motivation, engagement, and self-efficacy. The findings discussed below draw on both survey results and open-ended responses provided by students.

\textbf{RQ1 (Motivation and Engagement):} DeliverC was effective in capturing student interest through its game-based format. As shown in Figure~\ref{fig:unique_students}, 25 students participated in Level 1, showing strong initial engagement. While participation dropped to 20 students in Level 2 and 8 in Level 3, this decline likely reflects the optional, extra-credit setup and increased task difficulty, not a lack of interest. Survey responses (Figure~\ref{fig:survey}) support this interpretation: most students agreed that the game was more motivating than traditional practice. Some students described the experience in ways that suggest it was engaging and interactive, helping them better grasp the concept of pointers. This aligns with established constructs such as situational interest, engagement, and flow, which are known to support sustained attention and enjoyment in learning environments that offer challenge and interactivity. These findings are consistent with prior research suggesting that well-designed educational games can enhance motivation and make learning more engaging and interactive~\cite{yassine2017serious, kletenik2017serious, mathrani2016playit, videnovik2023game, zhan2022effectiveness}. 

\textbf{RQ2 (Learning Progress and Self‑Efficacy)}: The step-by-step design of DeliverC helped students gradually strengthen their understanding. Figure~\ref{fig:violin_attempts} shows that although only eight students reached Level 3, their average and median number of attempts dropped compared to Level 2. This suggests that working through the earlier levels helped them learn from their mistakes and make fewer errors later on. Figure~\ref{fig:survey} supports this pattern from a qualitative angle—more than 75\% of students agreed that the game helped them reflect on what they knew and feel more confident as they moved forward. Many students said the challenges pushed them to rethink what they had learned and apply it in new ways. 
These patterns reflect key constructs such as self-efficacy, metacognition, and mastery learning, which emphasize the value of reflection, confidence-building, and iterative practice in developing deeper understanding. Altogether, the results suggest that the scaffolded gameplay encouraged self-regulated learning by giving students repeated practice and building their confidence in solving pointer-related problems. This aligns with earlier studies showing that structured, interactive tasks and timely feedback can improve students’ sense of progress and confidence in programming~\cite{munshi2023analysing, lee2024harnessing, jurgensmeier2024generative, lyu2025will, zhan2022effectiveness, mathrani2016playit}.

\textbf{RQ3 (AI-Generated Feedback Effectiveness):} 
Many students found the AI feedback helpful, although a few noted that it could be improved. Some students reported that they adjusted their strategy in later tasks based on the suggestions they received, demonstrating that the system facilitated learning through trial and revision. This kind of interaction supports ideas such as formative feedback and self-regulated learning, which focus on providing timely help and promoting gradual improvement. Still, as Figure~\ref{fig:survey} shows, feedback clarity did not score as high as other areas. A few students mentioned in comments that some hints felt unclear or confusing. This is a common issue with AI-based tools—they can respond quickly and personally, but the explanations are not always straightforward to follow~\cite{akhoroz2025conversational, kazemitabaar2024codeaid, zhang2024students, mollick2024ai, lee2024harnessing}. Even so, students seemed to value the feedback overall, especially in terms of personalization and support. The lower ratings around clarity suggest that future versions of DeliverC should focus on making feedback more understandable.

Overall, our findings reinforce existing research on the value of adaptive, game-based environments in programming education, while also highlighting the need for clearer AI-generated feedback~\cite{munshi2023analysing, lee2024harnessing, mor2020game, videnovik2023game}. DeliverC adds to this body of work by demonstrating how personalized GenAI support can be effectively integrated into systems programming instruction—an area traditionally underserved by educational games.

\section{Threats to Validity}


This study was conducted in a single summer offering of a systems programming course with 37 students, of whom 25 chose to engage with the \textit{DeliverC} game. Since the activity was optional and tied to extra credit, the results may reflect self-selection bias: students who were already more motivated or confident may have been more likely to participate. Additionally, only a subset of students reached \texttt{Level 3}, which limits the generalizability of the findings at higher levels. While survey responses showed strong agreement across multiple educational constructs, they are based on self-reported perceptions, which may not always align with actual learning gains. To mitigate this, we triangulated findings with log data on task attempts and open-ended feedback. 
\section{Limitations and Future Work}



DeliverC was effective in helping students understand C pointers; however, its design is currently tied to a single programming topic in a specific course. This makes it difficult to apply the same game directly to other areas, especially outside programming, which raises concerns about scalability. Some students also mentioned that while the AI-generated hints were helpful overall, a few were unclear or lacked enough detail.

To improve the tool, future efforts will aim to extend DeliverC to cover more programming topics and test it in various programming courses. We also plan to refine the LLM feedback to make it more transparent and more useful, using prompt adjustments and student feedback. Ultimately, we aim to investigate adaptive challenge sequencing and evaluate long-term learning outcomes through longitudinal studies.
\section{Conclusion}




This study examined DeliverC, a GenAI-powered game designed to support the learning of C pointers in a programming course. Analysis of gameplay data and survey responses revealed that the tool effectively engaged students, encouraged reflective learning, and provided valuable AI-generated feedback. While some issues with clarity of feedback were noted, overall, student perceptions were positive. To our knowledge, DeliverC is one of the first tools to combine generative AI with game-based learning for teaching low-level programming concepts, such as pointers, thereby addressing a critical gap in CS education.

\bibliographystyle{ACM-Reference-Format}
\bibliography{SIGCSE-DeliverC}
\end{document}